\documentclass[twocolumn,english,preprintnumbers,amsmath,amssymb,pre,longbibliography]{revtex4-1}

\usepackage{graphicx}
\usepackage{color}

\begin{document}

\title{Pressure-Induced Mechanical Instabilities in Cubic SiC: Structural 
       and Electronic Properties}

\author{Carlos P. Herrero$^1$, Eduardo R. Hern\'andez$^1$, 
        Gabriela Herrero-Saboya$^2$, and Rafael Ram\'irez$^1$}

\affiliation{$^1$Instituto de Ciencia de Materiales de Madrid,
           Consejo Superior de Investigaciones Cient\'ificas (CSIC),
           Campus de Cantoblanco, 28049 Madrid, Spain  \\
         $^2$CNR-IOM Democritos National Simulation Center,
           Istituto Officina dei Materiali, c/o SISSA,
           via Bonomea 265, IT-34136 Trieste, Italy}	 

\date{\today}

\begin{abstract}
Silicon carbide is widely used in electronics, ceramics, 
and renewable energy due to its exceptional hardness and resistance. 
In this study, we investigate the effects of hydrostatic and uniaxial 
pressure (both compressive and tensile) on the structural and electronic 
properties of $3C$-SiC. Our analysis is based on atomistic molecular 
dynamics (MD) simulations using an efficient tight-binding Hamiltonian, 
whose accuracy is validated against density functional theory calculations.
Moreover, to account for nuclear quantum effects, we employ path-integral 
MD simulations. Our results show significant changes in the direct 
electronic gap as a function of temperature and pressure, with a 
renormalization of about 80~meV due to zero-point motion. Under 
hydrostatic tensile pressure, the direct band gap $E_{\Gamma}$ 
vanishes at the material's mechanical stability limit 
(spinodal point, where the bulk modulus $B \to 0$). For uniaxial pressure, 
we observe instabilities (Young's modulus $Y \to 0$) at approximately 
90~GPa for both tension and compression, where $E_{\Gamma} \to 0$.
Additionally, we analyze the pressure dependence of the internal energy, 
lattice parameter, and bond length, along with their finite-temperature 
fluctuations, which exhibit anomalies near the instability points.
\end{abstract}

\maketitle

\section{Introduction}

Silicon carbide is a versatile material renowned for its 
outstanding properties. It exhibits high hardness and excellent 
thermal conductivity, making it useful for electronic and power 
device applications. Additionally, SiC's electrical conductivity 
enhances its suitability for semiconductor technologies, particularly 
in high-temperature and high-power electronics \cite{sc-sh17,sc-pa22}.
With a high melting point and remarkable mechanical strength at 
elevated temperatures, it also plays an important role in industries 
such as aerospace and automotive \cite{sc-sh23,sc-wa21}.

The cubic $3C$-SiC phase is one of the most well-known forms of silicon 
carbide, stable under ambient conditions. Its mechanical properties 
have been extensively studied over the years through both theoretical 
\cite{sc-le82,sc-ra21,sc-pe22,sc-sh00,sc-le15b,sc-he23} 
and experimental methods 
\cite{sc-zh13,sc-ni17,sc-da17,sc-da18,sc-ki22}, owing to 
their significance in both fundamental research and technological 
applications.
In addition, the behavior of semiconducting materials, particularly 
silicon carbide, under high-pressure conditions has recently attracted 
increasing interest. This is largely due to the potential role of such 
materials in the extreme environments of carbon-rich exoplanets, where 
high-pressure conditions can significantly alter their physical 
properties \cite{sc-ni17,sc-ki22}.

Studying the behavior of materials under extreme pressure conditions 
involves not only large compressive pressure but also tensile stress 
(negative pressure). This aspect is particularly relevant for 
investigating unexplored metastability regions of solids 
\cite{he03b,sc-iy14,sc-pe15,sc-ni19,sc-ve22}.
In recent years, the experimentally accessible range of hydrostatic 
(or quasi-hydrostatic) tensile pressure has expanded, accompanied 
by a growing understanding of material properties in environments 
that are challenging to control in the laboratory 
\cite{sc-da10,sc-mo00,sc-du02,sc-ve22}.

Experimental studies of solids at negative pressure are 
relatively uncommon due to the metastable nature of these conditions, 
which are typically achievable only for brief duration. 
Notably, carbon-based materials have been investigated under tensile 
stress using ultrasonic cavitation and shock waves generated by 
picosecond laser pulses \cite{sc-ab14b,sc-kh08}. Additionally, 
low-density clathrates of C, Si, and Ge, metastable at ambient 
pressure, have been successfully synthesized \cite{sc-ba14,sc-gu06}.
In this context, silicon has been explored under negative pressure 
through molecular dynamics (MD) simulations, enabling the study of 
cavitation and crystal-liquid interfaces under such conditions 
\cite{sc-wi03,sc-da10b}. These simulations also revealed a transition 
to a clathrate phase at approximately 
$P \approx -2.5$~GPa \cite{sc-ka05}.

Various computational methods have been employed for research 
on $3C$-SiC, especially those based on density functional theory 
(DFT) at $T = 0$ \cite{sc-le15b,sc-ki17,sc-sh18,sc-ra21,sc-da22}. 
To account for finite-temperature properties, including anharmonic 
effects, classical atomistic simulations such as MD have been 
performed by several research groups 
\cite{sc-sh00,sc-ka05,sc-iv07, sc-ku17,sc-ka21}. 
Taking into account that the Debye temperature $\Theta_D$ of 
$3C$-SiC is clearly higher than room temperature 
($\Theta_D \approx$ 1100~K \cite{sc-zy96}), the combination of 
quantum nuclear motion (or phonon quantization) and anharmonicity 
of lattice vibrations may affect the material's properties up to 
relatively high temperatures.

Limitations associated with classical MD simulations may be 
overcome by computational techniques that explicitly account 
for nuclear quantum motion, such as Feynman path integrals. 
These methods have recently gained importance in investigating 
various physical properties of materials, including diamond 
\cite{ra06,br20}, silicon \cite{no96}, boron nitride 
\cite{ca16,br22}, and graphene \cite{br15,he16}. 
Nuclear quantum effects have been found to significantly influence 
electronic band gaps \cite{ra08} and the isotopic dependence 
of the lattice parameter in $3C$-SiC \cite{he09c}.

The electron-phonon interaction in tetrahedral semiconductors 
has traditionally been studied from a theoretical perspective 
using perturbation theory \cite{sc-zo92,sc-ki89}. 
An alternative approach to investigate the coupling between 
vibrational and electronic degrees of freedom in solids involves 
simultaneously applying a path-integral description for nuclear 
dynamics along with an electronic structure method, such as 
the tight-binding (TB) Hamiltonian employed here. This allows 
both atomic nuclei and electrons in the solid to be treated as 
quantum particles within the Born-Oppenheimer approximation, 
thereby directly incorporating phonon-phonon and electron-phonon 
interactions into the atomistic simulation.

In this paper, we analyze the metastability zone of cubic SiC 
for hydrostatic pressure $P < 0$ (tension), as well as the effects 
of tensile and compressive uniaxial pressure on its structural and 
electronic properties. We present outcomes of MD simulations 
performed with an interatomic potential based on an efficient 
TB Hamiltonian. The reliability of results under
far-from-equilibrium conditions is evaluated by 
comparison with DFT calculations 
carried out at $T = 0$. We assess nuclear quantum effects by 
comparing outcomes of path-integral molecular dynamics (PIMD) 
simulations with those yielded by classical MD using the same 
TB model. Our results indicate that quantum corrections manifest 
themselves in the structural and electronic properties of $3C$-SiC, 
being observable even at temperatures higher than 300~K.

The present work makes two key contributions: first, it extends 
previous studies on cubic SiC under hydrostatic pressure to include 
uniaxial pressure; second, it examines the effects of both hydrostatic 
and uniaxial pressure (tension and compression) on the electronic 
structure of this material, with a particular focus on changes 
induced in the direct band gap at the $\Gamma$ point. 
Special emphasis is given to the limits of mechanical stability 
of the material under tension and compression, where 
the direct band gap $E_{\Gamma}$ vanishes.

The paper is organized as follows: Sec.~II outlines the computational
methods employed, including molecular dynamics simulations (II.A), 
the tight-binding approach (II.B), and the DFT method (II.C). 
In Sec.~III, we present the results and discussion for the 
energy (III.A), crystal structure (III.B), bond length (III.C), 
elastic constants (III.D), and electronic gap (III.E).
The main findings are summarized in Sec.~IV.

\section{Computational method}

\subsection{Molecular dynamics}

We investigate the structural, elastic, and electronic properties
of $3C$-SiC using two types of MD simulations, 
enabling us to determine the equilibrium states of the system under 
various pressure and temperature conditions. 
First, we employ classical MD simulations, where Newton's equations 
of motion are solved numerically to trace atomic trajectories 
over time. Second, we use PIMD simulations, 
which explicitly account for the quantum nature of 
atomic nuclei at finite temperatures.
The primary computational distinction between the two approaches lies 
in the representation of atomic nuclei: in PIMD each nucleus is 
modeled as a set of $N_{\rm Tr}$ (Trotter number) beads, mimicking 
classical particles arranged in a ring polymer 
\cite{fe72,gi88,ce95,he14}. This creates a pseudoclassical system 
that accurately reproduces quantum properties. Comparing results 
from classical MD and PIMD simulations allows us to evaluate the 
impact of nuclear quantum effects on various properties. 
Notably, the classical limit is achieved in this formulation by 
setting $N_{\rm Tr} = 1$, where the ring polymers collapse into 
single particles.

Our simulations were conducted in the isothermal-isobaric ($NPT$) 
ensemble, utilizing an interatomic potential derived from a 
TB Hamiltonian, as described in Sec.~II.B. 
The simulation algorithms were based on established methods 
from the literature \cite{tu92,ma99}.
For PIMD simulations, staging coordinates were used to describe 
the bead positions within the ring polymers, and each staging 
coordinate was coupled to a chain of four Nos\'e-Hoover thermostats 
to maintain a constant temperature. Similarly, a chain of four 
thermostats was attached to the barostat, allowing for the necessary 
volume fluctuations to achieve the target pressure \cite{tu10,he14}.

The equations of motion were integrated using the reversible 
reference system propagator algorithm (RESPA), which enables the 
use of different time steps for integrating fast and slow degrees 
of freedom \cite{ma96}. For fast dynamical variables, such as bead 
interactions and thermostats, a time step of $\delta t = 0.25$ fs 
was used, while a larger time step of $\Delta t = 1$ fs was employed 
for the slower dynamics associated with interatomic forces. 
Additional details on this type of simulation can be found in 
\cite{tu10,he16}.

Most of our simulations, including both classical and PIMD, were 
performed on $2 \times 2 \times 2$ supercells of the face-centered 
cubic unit cell of $3C$-SiC ($N = 64$ atoms) with periodic boundary 
conditions. To verify the convergence of the results, additional 
simulations were conducted using $3 \times 3 \times 3$ supercells 
(216 atoms). Energy convergence was further assessed at $T = 300$~K 
for even larger cells, up to $N = 384$ atoms (see below).

The configuration space was sampled across a temperature range of 
50 to 1500~K. Some classical simulations were carried out at 
$T = 10$~K to study the stability limit under pressure without
the larger volume fluctuations appearing at higher $T$.
We considered hydrostatic pressures from $-44$~GPa 
(tension) to 60~GPa (compression), as well as uniaxial pressure $P_x$ 
applied along the [100] crystal axis, varying between $-84$ and 90~GPa. 
For both classical and quantum simulations, $2 \times 10^5$ steps 
were used for system equilibration, followed by $8 \times 10^6$ 
steps to compute average properties.
In PIMD simulations, the Trotter number $N_{\rm Tr}$ was 
temperature-dependent, following the relation 
$N_{\rm Tr} T = 6000$~K. This ensured a nearly constant 
accuracy across the temperature range \cite{he16}.

In the following, we will call $P$ and $P_x$ the hydrostatic and
uniaxial pressure, respectively. In the notation of elasticity, we have 
$\sigma_{xx} = \sigma_{yy} = \sigma_{zz} = -P$ in the first case,
and $\sigma_{xx} = -P_x$ in the second, where
$\{ \sigma_{ij} \}$ is the stress tensor.

\subsection{Tight-binding method}

To define the Born-Oppenheimer surface for nuclear motion in both 
classical and PIMD simulations, we have employed an effective 
tight-binding Hamiltonian \cite{po95}.
In principle, it could be possible to use {\em ab initio} 
methods (e.g., DFT) for finite-temperature PIMD simulations,
but in practice this procedure would considerably reduce the extent 
of simulation trajectories and/or the simulation-cell size.

In the present calculations, we use a method that considers the quantum
character of the electrons (TB Hamiltonian) for MD simulations
(classical atomic nuclei), complemented with a procedure to take into
account the quantum nature of atomic nuclei in PIMD simulations,
as described above in Sec.~II.A. 
This TB method uses a non-orthogonal Hamiltonian developed by 
Porezag {\it et al.} \cite{po95}, based on DFT calculations in the
local density approximation (LDA).
Specifically, the TB formalism used in our MD simulations was
adapted from the package TROCADERO \cite{si-ru03}.
The actual parameterization for structures including C and Si
atoms was given in Ref.~\cite{gu96}. Atomic orbitals are defined
as eigenfunctions of properly built pseudoatoms, with overlap 
matrices and Hamiltonian matrix elements tabulated as functions 
of the internuclear distance.
The non-orthogonality of the atomic basis provides a transferable
TB parametrization to complex systems \cite{po95}.

This TB parameterization has been employed to study bulk silicon 
carbide \cite{sc-me96,sc-be05}, isotopic and quantum effects in
$3C$-SiC \cite{he09c,sc-he24}, along with its surface reconstructions 
\cite{gu96}. In recent years, it was applied to analyze several 
aspects of newly synthesized SiC monolayers \cite{he22,sc-po23}.
The efficiency of TB methods to describe a collection of properties 
in condensed matter and molecular systems was reviewed
by Goringe {\it et al.} \cite{go97}.

In our classical MD simulations, atomic motion is governed by Newton's equations, 
with both total energy and interatomic forces derived from the TB Hamiltonian. 
This approach enables tracking of the material’s evolving electronic structure 
throughout the simulation.
A similar methodology is applied in PIMD simulations, where interatomic forces 
are also computed using the TB Hamiltonian. In this context, the electronic
structure is obtained by averaging over imaginary time, specifically across
the $N_{\rm Tr}$ time slices (beads).

In this work, the electronic degrees of freedom in reciprocal 
space have been sampled by considering only the $\Gamma$ point 
(${\bf k} = 0$). We verified that introducing additional 
${\bf k}$ points results in a minor change in the total energy, 
with no significant impact on the energy differences
that are crucial to our analysis.
Although the minimum-energy $E_0$ (the classical limit for $T = 0$) 
exhibits a slight shift, this shift diminishes as the simulation 
cell is enlarged.

In Fig.~1 we present the energy $E - E_0$ of unstressed $3C$-SiC 
($P = 0$) derived from our simulations at $T = 300$~K.
Solid circles and open squares represent results of classical MD 
and PIMD simulations, respectively.
In both classical and quantum cases, $E - E_0$ converges to
78 and 130 meV/atom, respectively. For $N > 50$, the shift 
in $E - E_0$ due to the system size is less than 1~meV/atom.

\begin{figure}
\vspace{-7mm}
\includegraphics[width= 7cm]{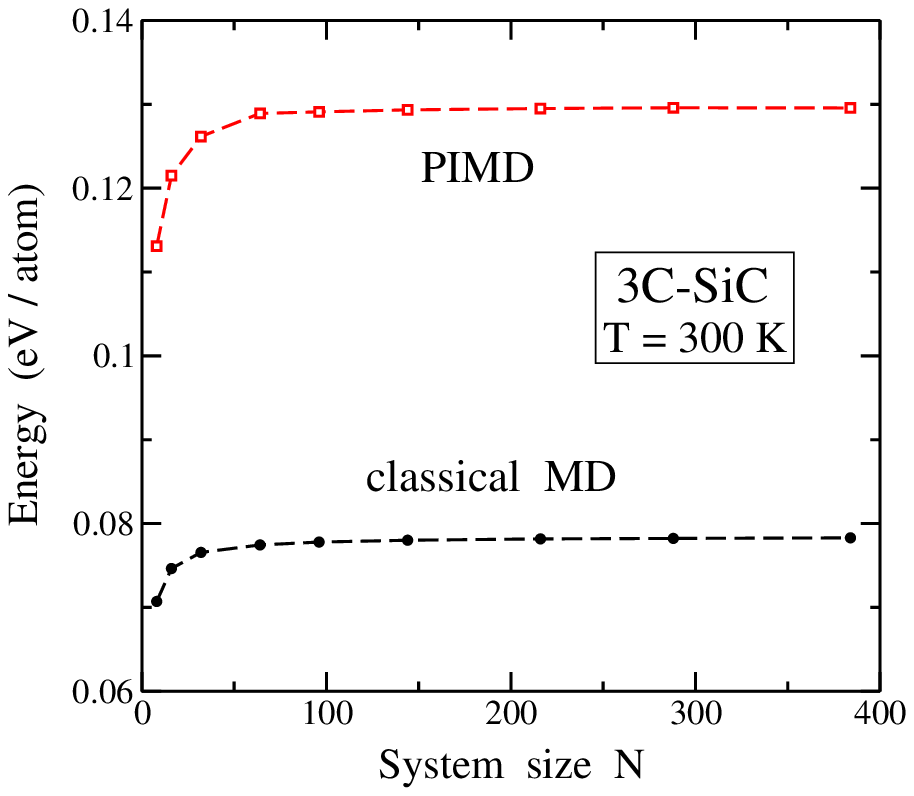}
\vspace{-5mm}
\caption{Energy as a function of system size for $3C$-SiC,
obtained from MD simulations with the TB Hamiltonian employed here.
Data points are energy results for classical (solid cricles) and
PIMD simulations (open squares) at $T$ = 300~K.
The energy reference corresponds to the minimum energy value for
each size $N$.
Lines are guides to the eye.
}
\label{f1}
\end{figure}

\subsection{DFT calculations}

We evaluated the precision of the considered tight-binding  method 
for predicting the structural and electronic properties of $3C$-SiC by 
comparing its results with DFT calculations for 
this material at $T = 0$. To achieve this, electronic structure 
calculations were performed using the Quantum ESPRESSO code 
\cite{sc-gi09,sc-gi17}.

We utilized the Perdew-Burke-Ernzerhof exchange-correlation functional 
optimized for solid-state systems (PBEsol) \cite{sc-pe08}, in 
combination with projector-augmented-wave (PAW) pseudopotentials for 
Si and C atoms \cite{sc-ps23}. 
For the plane-wave basis set, we employed cutoff values of 400 Ry for 
the charge density and 45 Ry for the kinetic energy. The calculations 
were performed on a cubic unit cell of SiC containing eight atoms, with 
periodic boundary conditions. Reciprocal space integration was carried 
out using a $10 \times 10 \times 10$ Monkhorst-Pack grid \cite{sc-mo76}.

DFT calculations have previously been employed to investigate several
properties of $3C$-SiC, including its structural, mechanical, 
lattice-dynamical, thermodynamic, and electronic characteristics 
\cite{sc-pa94b,sc-ka94,sc-ka94b,sc-ca20}. 
This kind of calculations have significantly contributed to advancing 
the understanding of the silicon carbide phase diagram, particularly 
with respect to high-pressure phase transitions 
\cite{sc-le15b,sc-ki17,sc-sh18,sc-ra21}.

\section{Results and discussion}

\subsection{Energy}

In this section, we examine the energy of $3C$-SiC, derived from 
classical MD and PIMD simulations performed in the $NPT$ ensemble 
across a range of pressures and temperatures. At a given external 
pressure $P$, the energy at $T = 0$ can be expressed as 
$E = E_{\rm cl}^0 + E_{\rm ZP}$, where $E_{\rm cl}^0$ represents 
the classical energy of static atoms, and $E_{\rm ZP}$ denotes 
the zero-point energy. Both quantities are normalized as energy
per atom.

\begin{figure}
\vspace{-7mm}
\includegraphics[width= 7cm]{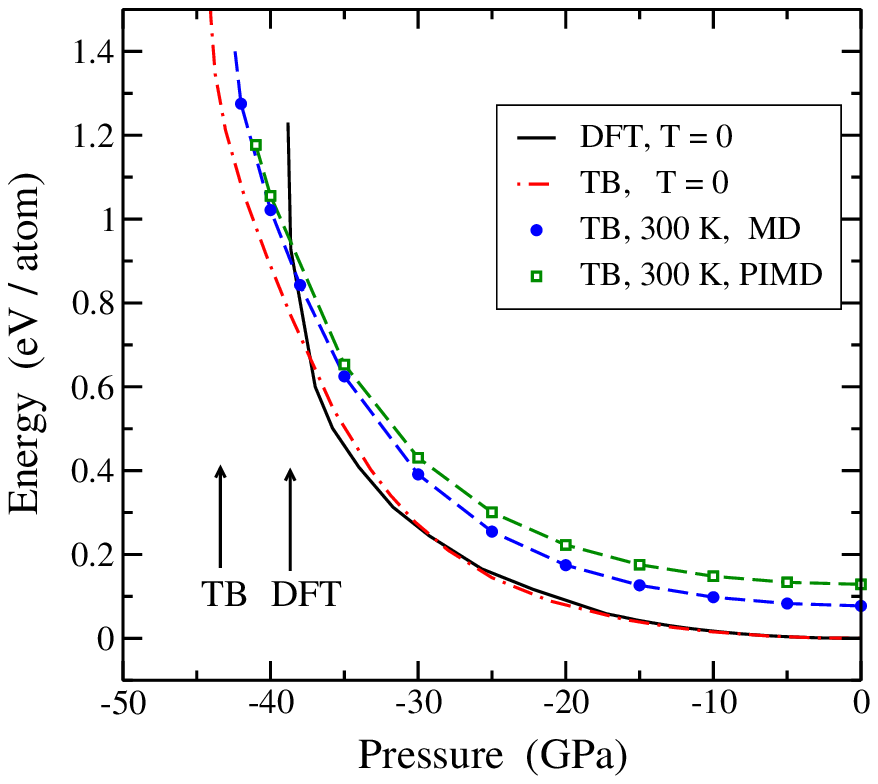}
\vspace{-5mm}
\caption{Energy per atom as a function of hydrostatic pressure.
The solid and dashed-dotted lines represent results derived
from DFT and TB calculations, respectively, at $T = 0$.
Symbols correspond to results from classical MD (solid circles)
and PIMD simulations (open squares) at $T = 300$~K.
Dashed lines are guides to the eye.
Two vertical arrows indicate the spinodal pressure corresponding
to DFT and TB models at $T = 0$.
}
\label{f2}
\end{figure}

We begin by analyzing the dependence of the classical energy, 
$E_{\rm cl} - E_0$, on tensile hydrostatic pressure ($P < 0$), 
as shown in Fig.~2. The solid and dashed-dotted lines represent 
the DFT and TB results at $T = 0$, respectively.
Our analysis shows that the TB curve closely tracks the DFT 
calculations up to a pressure of approximately $-35$~GPa. Beyond 
this point, the DFT data exhibit a noticeable increase, eventually 
leading to a mechanical instability in the crystal structure at 
$P_s = -39(1)$~GPa. At this point, the pressure derivative of the 
energy diverges, $\partial E / \partial P \to -\infty$.
Similarly, for the TB results at $T = 0$, we observe a divergence 
at $P_s = -44$~GPa. In both cases, the spinodal pressure, $P_s$, 
is marked by a vertical arrow in Fig.~2.

For $P > 0$, the energy increases with rising compression, 
exhibiting no anomalies within the range where cubic SiC remains 
the stable phase ($P \leq 60$~GPa). This behavior has been 
well-documented in several previous studies 
\cite{sc-le15b,sc-ki17,sc-sh18,sc-ra21,sc-he24}, and thus will 
not be elaborated upon here.

In a quantum framework, the zero-point energy per atom, $E_{\rm ZP}$, 
can be calculated in a harmonic approximation from the vibrational 
density of states (VDOS), $\rho(\omega)$, as:
\begin{equation}
  E_{\rm ZP} = \frac{1}{2} \int_0^{\omega_m} \frac{\hbar \omega}{2} \, 
	\rho(\omega) \, d\omega \; ,
\label{ezp}
\end{equation}
where $\omega_m$ represents the maximum frequency in the phonon 
spectrum. The prefactor $1/2$ in Eq.~(\ref{ezp}) ensures the energy 
is calculated per atom, as $\rho(\omega)$ is normalized to account 
for the six degrees of freedom (corresponding to one Si and one C 
atom) in a crystallographic unit cell:
\begin{equation}
        \int_0^{\omega_m} \rho(\omega) \, d\omega = 6 \; .
\label{intg}
\end{equation}
The mean vibrational frequency, $\overline{\omega}$, is defined as:
\begin{equation}
   \overline{\omega} = \frac{1}{6} \int_0^{\omega_m} \omega \, 
	  \rho(\omega) \, d\omega \; ,
\label{ommed}
\end{equation}
which gives: $E_{\rm ZP} = 3 \hbar \, \overline{\omega} / 2$.


\begin{table*}[ht]
\caption{Energy of $3C$-SiC derived from classical MD ($E_{\rm cl}$)
and quantum PIMD simulations ($E_{\rm q}$) at $T$ = 0, 300 and 1000~K
under hydrostatic pressure $P$.
Zero-temperature data are extrapolations of finite-$T$ results.
The difference between quantum and classical data is denoted
$\Delta E$. Energies are given in meV/atom.
}
\vspace{0.6cm}

\begingroup
\setlength{\tabcolsep}{10pt}
\renewcommand{\arraystretch}{1.5}
\centering
\setlength{\tabcolsep}{10pt}
\begin{tabular}{|c|c c c| c c c| c c c|}
\cline{2-10}
\multicolumn{1}{c}{} &
      \multicolumn{3}{|c|}{$P = 0$} &
      \multicolumn{3}{c|}{$P = -30$~GPa} &
      \multicolumn{3}{c|}{$P = 60$~GPa} \\ [2mm]
\cline{1-10}
  $T$ (K) & $E_{\rm cl}$ & $E_{\rm q}$ & $\Delta$ E & $E_{\rm cl}$ & $E_{\rm q}$
        & $\Delta$ E & $E_{\rm cl}$ & $E_{\rm q}$ & $\Delta$ E
  \\[2mm]
\hline
     0 &   0 & 112 & 112 & 291 & 399 & 108 & 303 & 422 & 119 \\
   300 &  77 & 129 &  52 & 391 & 431 &  40 & 378 & 436 &  58 \\
  1000 & 264 & 281 &  17 & 621 & 633 &  12 & 547 & 566 &  19 \\
\hline
\end{tabular}
\endgroup
\label{ener_simul}
\end{table*}

The VDOS of $3C$-SiC for the present 
TB model was calculated and reported in Ref.~\cite{sc-he24}.
For the unstressed material ($P = 0$), it yields a zero-point 
energy of $E_{\rm ZP} = 115$~meV/atom.
By extrapolating the energy $E$ obtained from our PIMD simulations to 
$T = 0$, we determine $E_{\rm ZP}$ values of 108, 112, and 119~meV/atom 
for silicon carbide under hydrostatic pressures of $P = -30$, 0, and 
60~GPa, respectively. These values correspond to mean frequencies of 
$\overline{\omega} = 581$, 602, and 640~cm$^{-1}$, respectively, 
following the expected trend under tensile and compressive pressure 
(see Table~I).
At $T = 0$, the quantum energy $E_{\rm q}$ lies between the classical 
and quantum results at $T = 300$~K shown in Fig.~2 and is omitted from 
the figure for clarity.
Notably, the average frequency $\overline{\omega}$ obtained from PIMD 
simulations for $P = 0$ is 3~meV lower than that derived from the VDOS 
in the harmonic approximation. This shift in the zero-point energy 
(approximately 3\%) arises primarily as an effect of 
anharmonicity in the kinetic energy at $T = 0$, as demonstrated 
by perturbation theory calculations \cite{sc-he24}.

Turning to the results of our simulations at finite temperatures, 
Fig.~2 shows the energy at $T = 300$~K, with solid circles 
representing classical MD results and open squares denoting 
PIMD data.
At low pressures ($|P| \lesssim 5$~GPa), the classical results exhibit 
a nearly constant increase of $3 k_B T$ (78~meV/atom) compared 
to the data at $T = 0$, consistent with predictions from a harmonic 
model of lattice vibrations.
As tensile pressure increases, the difference between the 
classical energies at $T = 300$~K and $T = 0$ grows, reaching 
100~meV/atom at $P = -30$~GPa. This behavior suggests the emergence 
of anharmonicity in the SiC crystal under conditions far from 
equilibrium ($P = 0$).

Examining the results of the PIMD simulations, we observe that 
for $P = 0$, the energy shows an increase of 52~meV/atom relative 
to the classical energy, yielding a total of 129~meV/atom at 300~K.
Notably, this energy increase due to quantum nuclear motion 
varies from 40~meV/atom under tensile pressure of $-30$~GPa to 
58~meV/atom under compressive pressure of 60~GPa (see Table~I).
At 300~K, both the classical and quantum results exhibit a 
divergence in the slope of the energy-pressure curve for
$P \approx -43$~GPa.

As temperature increases, the results of classical MD and PIMD 
simulations converge, as nuclear quantum effects diminish in 
significance. For example, at $T = 1000$~K and $P = 0$, the 
difference between the two datasets is 17~meV/atom. 
This difference decreases further to 12~meV/atom under a tensile 
pressure of $-30$~GPa (see Table~I).
As the mean vibrational frequency $\overline{\omega}$ increases 
with rising $P$, the derivative $\partial E_{\rm ZP} / \partial P$ 
is positive.  This results in a rise of the energy difference 
$\Delta E = E_{\rm q} - E_{\rm cl}$ with increasing $P$, 
which is most pronounced at low temperatures.

The zero-point energy $E_{\rm ZP}$ derived from our PIMD simulations
can be connected to the mean frequency $\overline{\omega}$, as
indicated above, which in turn is related to the Debye frequency
$\omega_D = k_B \Theta_D / \hbar$, yielding (see Appendix~A):
\begin{equation}
    \omega_D = \frac89 \, \frac{E_{\rm ZP}}{\hbar} 
	     =  \frac43 \, \overline{\omega}   \; .
\label{omd}
\end{equation}
For $P = 0$, the zero-point energy obtained from our PIMD
simulations gives a Debye temperature $\Theta_D =$ 1155~K,
near values presented in the literature for $3C$-SiC:
1123~K derived from specific heat measurements \cite{sc-zy96}
and 1147~K obtained from {\em ab initio} calculations combined
with a quasi-harmonic approach \cite{sc-xu18}.

\subsection{Pressure effects on the crystal structure}

For unstressed $3C$-SiC, the lattice parameter corresponding to 
the minimum-energy configuration calculated using the TB Hamiltonian 
is $a_0 = 4.348$~\AA, yielding a crystal volume of 
$V_0 = 10.277$~\AA$^3$/atom. Similarly, our DFT calculations provide 
$a_0 = 4.358$~\AA, which corresponds to a volume of 
$V_0 = 10.346$~\AA$^3$/atom, consistent with previous {\em ab initio} 
results \cite{sc-le15b}.
From the outcome of TB-PIMD simulations at $P = 0$, we obtain 
a low-temperature extrapolated lattice parameter value of 
$a_{\rm min} = 4.359$~\AA. This indicates a zero-point expansion 
arising from quantum nuclear motion, quantified as 
$\delta a = 1.1 \times 10^{-2}$~\AA\ \cite{sc-he24}.
Furthermore, the lattice parameter $a$ derived from our 
finite-temperature PIMD simulations is in good agreement with 
values determined for $3C$-SiC via x-ray diffraction experiments: 
$a = 4.36$~\AA\ at 297~K and atmospheric pressure \cite{sc-sl75,ra08}.

\subsubsection{Hydrostatic pressure}

We now examine the effect of hydrostatic pressure $P$ on the 
crystal volume. Fig.~3 illustrates the pressure dependence of the 
lattice parameter $a$ at $T = 0$, as obtained from TB calculations 
(dashed-dotted line) and DFT calculations (solid curve).
The two curves exhibit strong agreement across a wide range of 
pressures, encompassing both tensile and compressive stress, 
as shown in the figure. Notably, the slope of both curves 
increases as the tensile stress is raised.
For $P < 0$, the lattice parameter approaches a mechanical 
instability under the same tensile stresses indicated in Fig.~2 
for the energy at $T = 0$ (marked by vertical arrows).

\begin{figure}
\vspace{-7mm}
\includegraphics[width= 7cm]{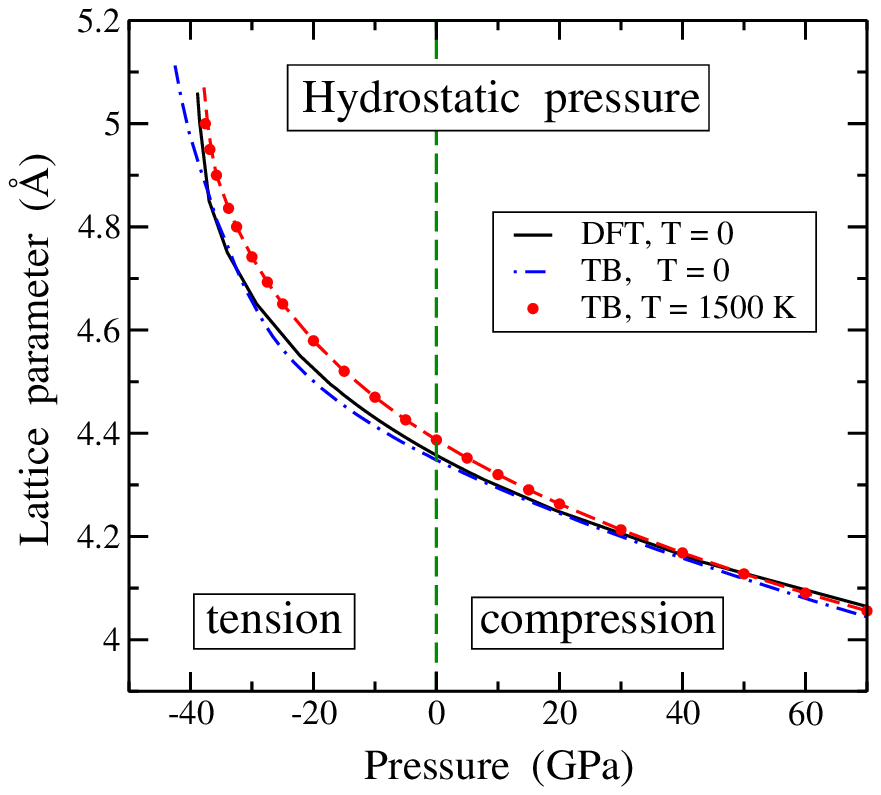}
\vspace{-5mm}
\caption{Lattice parameter vs hydrostatic pressure.
The solid and dashed-dotted lines indicate results derived
from DFT and TB calculations, respectively, at $T = 0$.
Solid circles represent results from classical MD
simulations at $T = 1500$~K.
The dashed line through the data points is a guide to the eye.
}
\label{f3}
\end{figure}

Fig.~3 also includes results from classical MD simulations conducted 
at high temperature ($T = 1500$~K, shown as solid symbols). These 
results reveal a significant lattice expansion under tensile stress, 
with a value of $\delta a = 8.4 \times 10^{-2}$~\AA\ at $P = -30$~GPa 
relative to the zero-temperature result.
In contrast, under compression ($P > 0$), the lattice expansion is 
considerably smaller. For instance, at $P = 50$~GPa, the lattice 
expansion is reduced to $\delta a = 9 \times 10^{-3}$~\AA.

When a material with a given crystal structure expands,
it reaches a volume $V_s$ where mechanical stability is lost.
This corresponds to the spinodal point at the considered
temperature $T$, given by the condition:
\begin{equation}
 \left. \frac{\partial^2 F}{\partial V^2} \right|_{V_s} = 0 \; ,
\label{vsv}
\end{equation}
where $F$ is the Helmholtz free energy \cite{ca85}.
This condition is equivalent to the vanishing of the bulk modulus,
$B = - V \partial P / \partial V$, as 
$P = - \partial F / \partial V$.
Near a spinodal point, the dependence of the volume $V$ on $P$
is given by \cite{sp82,sc-he23}:
\begin{equation}
    V_s - V = c \, (P - P_s)^{1/2}  \; ,
\label{vsv2}
\end{equation}
where $P_s$ is the spinodal pressure at temperature $T$, 
and $c$ is a constant. For a cubic solid, putting $V = a^3$, 
we have for $a$ near the spinodal lattice parameter 
$a_s$: $a^3 - a_s^3 \approx 3 a_s^2 (a - a_s)$, so that
\begin{equation}
    a_s - a = \frac{c}{3 a_s^2} \, (P - P_s)^{1/2}  \; .
\label{asa}
\end{equation}
Fitting the data presented in Fig.~3 to Eq.~(\ref{asa}), 
yields $P_s = -44(1)$ and $-38(1)$~GPa for TB and DFT data at 
$T = 0$, respectively, in agreement with earlier estimations
\cite{sc-he23}. Note the shift of $P_s$ for the TB data at
1500~K, which turns out to be close to the DFT result
at $T = 0$.

At this point, we clarify that our use of the term “spinodal” does not 
refer to phase separation in the classical thermodynamic sense 
(such as in a binary mixture), driven by composition fluctuations. 
Instead, it denotes the mechanical stability 
limit associated with the loss of convexity of the free energy with 
respect to strain or volume. Within this context, the spinodal corresponds 
to the point at which the elastic moduli, or the relevant components of 
the stiffness tensor (in particular, the bulk modulus $B$) vanish, 
signaling the onset of mechanical instability.

Our study thus focuses specifically on this second type of instability, 
which involves variations in specific volume (or equivalently, lattice 
strain) under pressure. We do not suggest the existence of a second 
thermodynamic phase as implied by classical spinodal decomposition. 
Rather, we identify a mechanical instability that can lead to structural 
transformations, amorphization, or fracture, depending on the deformation 
pathway and kinetic conditions.

Put differently, at the spinodal point, the crystal lattice can no longer 
sustain uniform tension and becomes unstable with respect to infinitesimal 
strain fluctuations. During our atomistic simulations near the spinodal 
pressure $P_s$, we observe large volume fluctuations that may precipitate 
the breakdown of the crystal structure. This breakdown is characterized 
by the sudden emergence of point and extended defects, potentially leading 
to fracture of the material.

\subsubsection{Uniaxial pressure}

\begin{figure}
\vspace{-7mm}
\includegraphics[width=7cm]{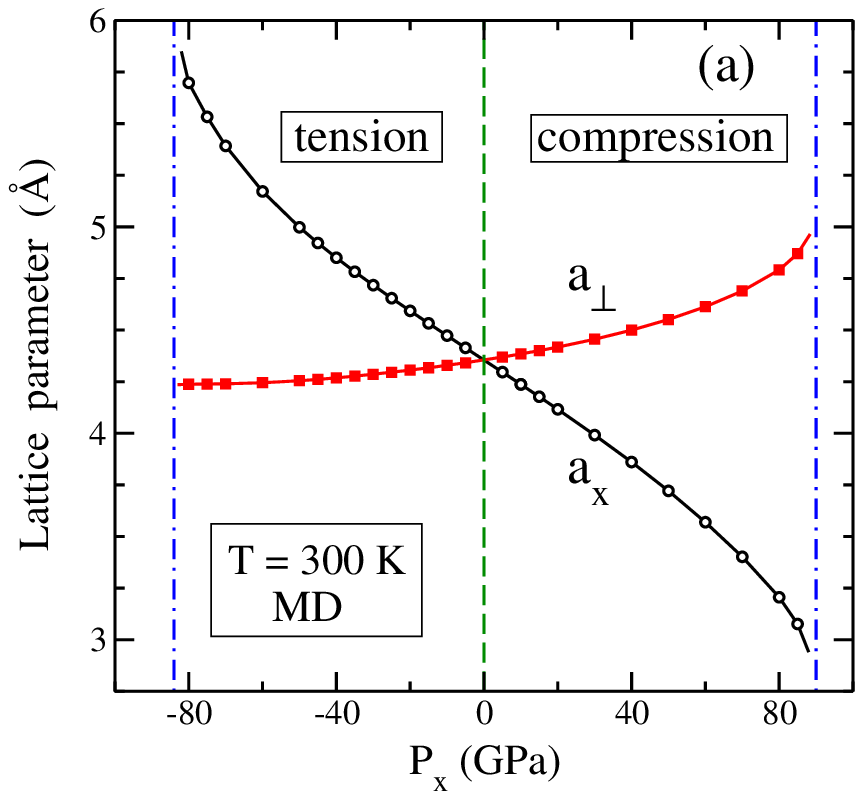}
\includegraphics[width=7cm]{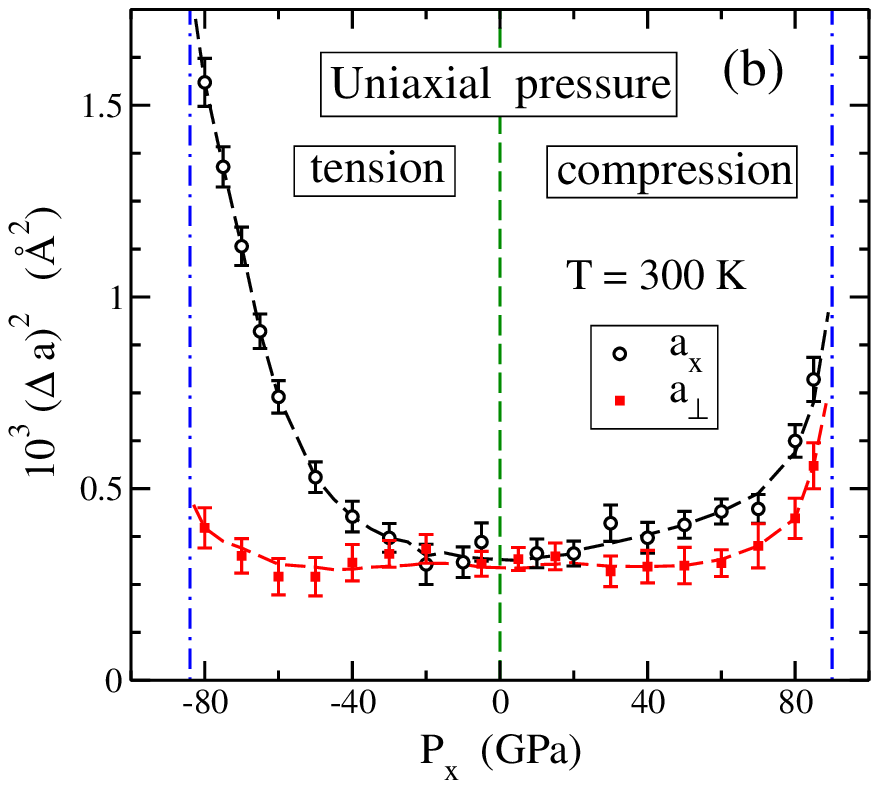}
\vspace{-5mm}
\caption{(a) Lattice parameter vs uniaxial pressure $P_x$.
Symbols represent results of classical MD simulations
at $T = 300$~K: open circles for $a_x$ and solid squares
for $a_{\perp}$ ($\perp = y, z$).
(b) Mean-square fluctuation of the lattice parameters.
Open and solid symbols are data points for $a_x$ and
$a_{\perp}$, respectively. Lines drawn through the data points
are guides to the eye.
Dashed-dotted lines show the instability pressures
$P_x^{i1} = -84$~GPa and $P_x^{i2} = 90$~GPa.
}
\label{f4}
\end{figure}

We now turn to the effects of uniaxial pressure $P_x$ on the 
structural properties. Fig.~4(a) illustrates the dependence of the 
lattice parameters $a_x$ (parallel to the applied pressure) and 
$a_{\perp}$ ($\perp = y, z$, perpendicular to the applied pressure) 
on $P_x$.  Open circles represent data for $a_x$, while solid 
squares correspond to $a_{\perp}$.
These results were obtained from classical MD simulations. 
For simplicity, data derived from PIMD at 300~K are not shown, 
as they closely align with the classical results on the scale 
of the figure.
The parameter $a_x$ satisfies $\partial a_x / \partial P_x < 0$, 
with the slope steepening significantly under both tensile 
($P_x \approx -80$~GPa) and compressive ($P_x \approx 80$~GPa) 
pressure. This behavior suggests the emergence of mechanical 
instabilities in $3C$-SiC under extreme uniaxial pressures, 
as further discussed below.

The Young's modulus, $Y$, is a mechanical property that quantifies 
a material’s resistance to tensile or compressive deformation when 
subjected to an applied uniaxial pressure (in this case, $P_x$).
In the linear elastic regime, it is defined as the ratio between the 
stress $\sigma_{xx} = -P_x$ and the resulting axial strain 
$\epsilon_{xx} = \Delta a_x / a_x$.
Beyond the linear region, for uniaxial pressure applied along the 
[100] crystal axis, $Y$ can be expressed as:
\begin{equation}
    Y = -a_x \frac{\partial P_x}{\partial a_x} =
           - \frac{\partial P_x}{\partial \ln a_x}  \; .
\label{young}
\end{equation}
From our results at $T = 300$~K, we find a Young's modulus of 
$Y = 369$~GPa for $P_x = 0$, which coincides with the value
derived from stiffness elastic constants $C_{11}$ and $C_{12}$
\cite{sc-he23} through the expression \cite{sc-ho12}:
\begin{equation}
	Y = \frac{ (C_{11} + 2 C_{12}) (C_{11} - C_{12}) }
	  { C_{11} + C_{12} }  \; .
\label{young2}
\end{equation}
This result is in good agreement with the data reported in 
Ref.~\cite{sc-ra21}.

As shown in Fig.~4(a), $| \partial a_x / \partial P_x |$ increases 
sharply near a tensile pressure of $-80$~GPa and a compressive 
one of 80~GPa, approaching the limit $Y \to 0$ in both cases. 
This behavior indicates the onset of mechanical instabilities 
under these stress conditions.
By extrapolating the MD results, we estimate the instability
pressures to be $P_x^{i1} = -84(2)$~GPa for tension and 
$P_x^{i2} = 90(2)$~GPa for compression.

For the parameter $a_{\perp}$, we find 
$\partial a_{\perp} / \partial P_x > 0$, 
consistent with a positive Poisson's ratio, $\nu$, for this solid 
\cite{sc-ra21,sc-he24}. This ratio can be calculated as 
$\nu = -\Delta a_{\perp} / \Delta a_x$ for small values of $P_x$. 
From our calculations, we obtain $\nu = 0.23$, which agrees well 
with the value derived from the elastic constants of cubic SiC 
\cite{sc-he24}.
The derivative $\partial a_{\perp} / \partial P_x$ decreases under 
tensile pressures, and the $P_x$--$a_{\perp}$ curve becomes nearly flat 
for large $| P_x |$.  Consequently, 
$\partial a_{\perp} / \partial P_x \to 0$, and the effective 
Poisson's ratio approaches zero near the instability at $P_x^{i1}$.

Additional insights into the behavior of the solid near the 
instability points can be gained by analyzing the fluctuations 
of the lattice parameters during simulation runs. These results 
are presented in Fig.~4(b), where the mean-square fluctuations (MSFs) 
$(\Delta a_x)^2$ and $(\Delta a_{\perp})^2$ are shown as open circles 
and closed squares, respectively. The data were obtained from 
our classical MD simulations at 300~K.
A sharp increase in $(\Delta a_x)^2$ is observed near the 
instability points, both under tensile and compressive uniaxial pressure.
While the increase is less pronounced for $(\Delta a_{\perp})^2$, 
it remains noticeable in these regions.

The results shown in Fig.~4(b) correspond to classical MD simulations. 
For PIMD, we find that the MSFs $(\Delta a_x)^2$ and 
$(\Delta a_{\perp})^2$ are comparable to those presented in the 
figure.  However, this similarity does not extend to the fluctuations 
of the interatomic distances, which exhibit distinct behavior, 
as discussed below.

\subsection{Bond length}

For the minimum-energy state of $3C$-SiC, the TB Hamiltonian predicts 
a bond distance of $d^0_{\rm Si-C} = 1.883$~\AA, which closely 
agrees with the DFT result of 1.887~\AA. When accounting for 
zero-point nuclear motion, we obtain an adjusted bond distance of 
$d^0_{\rm Si-C} = 1.888$~\AA, derived by extrapolating 
finite-temperature TB-PIMD results to $T = 0$.

Classical simulations at finite temperature $T$ show that the bond 
distance $d_{\rm Si-C}$ exhibits a linear dependence up to 
$T \approx 1000$~K, following $d_{\rm Si-C} = d^0_{\rm Si-C} + b_1 T$
with $b_1 = 1.6 \times 10^{-5}$~\AA/K. This thermal expansion of 
the bond length is amplified under tensile pressure: at $P = -30$~GPa,
the slope increases to $b_1 = 4.4 \times 10^{-5}$~\AA/K, nearly three
times its value at $P = 0$. Conversely, under compression, the slope 
decreases; for $P = 60$~GPa, we obtain $b_1 = 5.2 \times 10^{-6}$~\AA/K,
approximately half of the unstressed value.
At low temperatures, the thermal expansion of the real solid deviates 
from this linear trend, as $\partial d_{\rm Si-C} / \partial T \to 0$
for $T \to 0$, in accordance with the third law of thermodynamics. 
This effect is confirmed by PIMD simulations \cite{sc-he24,ra08}, 
which account for zero-point expansion 
of both the atomic bond and the crystal lattice, as discussed earlier.

At a given temperature (e.g., $T = 300$~K), the Si--C bond length 
under hydrostatic pressure follows a similar trend to the lattice 
parameter (see Fig.~3), in the whole region where the solid is
mechanically stable. In particular, under tensile stress, 
$d_{\rm Si-C}$ increases rapidly near the spinodal pressure 
$P_s$. In contrast, under compression, no significant anomalies 
are observed, with $d_{\rm Si-C}$ decreasing smoothly across 
the stability range of the $3C$-SiC phase, as expected. 
At $P = 0$, the pressure derivative of the bond length is 
$\partial d_{\rm Si-C} / \partial P = -2.7 \times 10^{-3}$~\AA/GPa.

\begin{figure}
\vspace{-7mm}
\includegraphics[width=7cm]{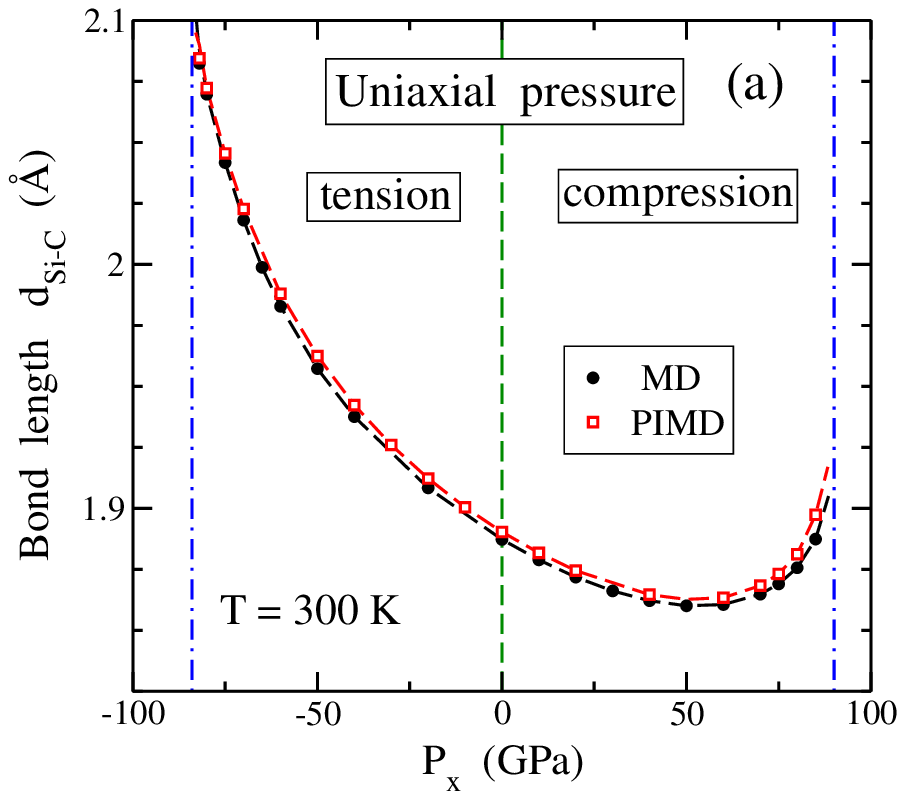}
\includegraphics[width=7cm]{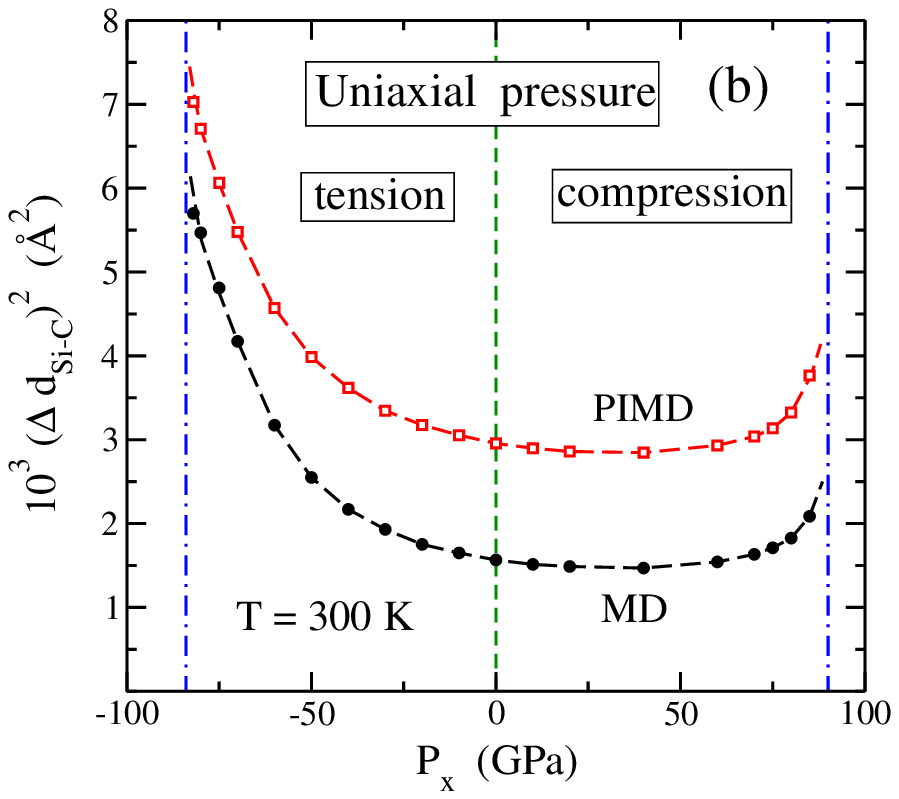}
\vspace{-5mm}
\caption{(a) Mean interatomic distance, $d_{\rm Si-C}$, as a
function of uniaxial pressure $P_x$, derived from classical
MD (solid circles) and PIMD simulations (open squares) at
$T$ = 300~K.  Error bars are less than the symbol size.
(b) Mean-square fluctuation of the interatomic distance,
$(\Delta d_{\rm Si-C})^2$ vs $P_x$. Solid circles:
classical MD; open squares: PIMD.
Error bars are in the order of the symbol size.
Lines connecting the data points are guides to the eye.
}
\label{f5}
\end{figure}

The dependence of the bond length, $d_{\rm Si-C}$, on uniaxial pressure 
$P_x$ is influenced by the crystal structure of $3C$-SiC, in which 
all Si--C bonds form the same angle, $\varphi = 54.7^{\circ}$, with 
the [100] axis. As a result, the effect of $P_x$ is uniform across 
all bonds. However, the response of $d_{\rm Si-C}$ to uniaxial pressure 
differs significantly from its behavior under hydrostatic pressure. 
This distinction is illustrated in Fig.~5(a), which shows the mean 
Si--C bond length as a function of $P_x$ at a temperature of 300~K. 
Symbols represent the results from classical MD (solid circles) 
and PIMD simulations (open squares), with the quantum 
results lying slightly above the classical ones on the scale of 
the figure.

The slope of the $P_x$--$d_{\rm Si-C}$ curve increases for growing
tensile stress, eventually diverging to $-\infty$ at the instability 
point $P_x^{i1} = -84$~GPa. At zero pressure, the derivative is
$\partial d_{\rm Si-C} / \partial P_x = -9.2 \times 10^{-4}$~\AA/GPa,
which is one-third of $\partial d_{\rm Si-C} / \partial P$ for 
hydrostatic pressure, 
as expected for uniaxial pressure in three-dimensional space. 
Under compression, the Si--C bond length reaches a minimum at
$P_x \approx 50$~GPa. For higher $P_x$, $d_{\rm Si-C}$
increases until the material undergoes mechanical instability 
at a compressive pressure of $P_x^{i2} \approx 90$~GPa.

Fig.~5(b) presents the MSF $(\Delta d_{\rm Si-C})^2$ obtained from 
classical (open circles) and quantum (open squares) simulations 
at 300~K. For the unstressed solid, the MSF from PIMD is 1.9 times 
larger than that from classical simulations. This ratio remains 
nearly constant for $P_x > 0$ but decreases under tensile pressure, 
reaching approximately 1.2 near the instability point at 
$P_x^{i1} = -84$~GPa. In both classical and quantum cases, 
the MSF exhibits a minimum at $P_x \approx 30$~GPa, and increases 
rapidly as the instability pressure is approached under both tension 
and compression. This increase is more pronounced under tensile 
pressure, mirroring the behavior of the bond length $d_{\rm Si-C}$
shown in Fig.~5(a).

\subsection{Elastic constants}

In the context of the mechanical stability of cubic silicon carbide, 
it is important to emphasize that its elastic constants undergo significant 
modifications within the pressure range considered in this study. 
Within the framework of the atomistic simulations presented here, these 
constants can be determined at finite temperatures by analyzing the strain 
response of the crystal to various applied stress conditions 
(i.e., different configurations of the stress tensor). 
Accordingly, we have computed the elastic stiffness constants of $3C$-SiC 
from MD simulations, following the methodology 
detailed in Ref.~\cite{sc-he23}.

At $T = 300$~K, classical MD simulations yield $C_{11} = 435(1)$~GPa and 
$C_{12} = 129(1)$~GPa. Compared to the corresponding values at $T = 0$~K, 
obtained using the same TB method, these results reflect a thermal-induced 
reduction of approximately 4\% and 8\%, respectively \cite{sc-he23}. 
When nuclear quantum effects are included, the elastic response further 
softens: PIMD simulations at 300~K give $C_{11} = 425(2)$~GPa and 
$C_{12} = 123(2)$~GPa \cite{sc-he24}. Although the elastic constant 
$C_{44}$ is less central to the present analysis, we note for completeness 
that our classical MD and PIMD simulations at 300~K yield values of 
238(1) and 235(1)~GPa, respectively.

\begin{figure}
\vspace{-7mm}
\includegraphics[width= 7cm]{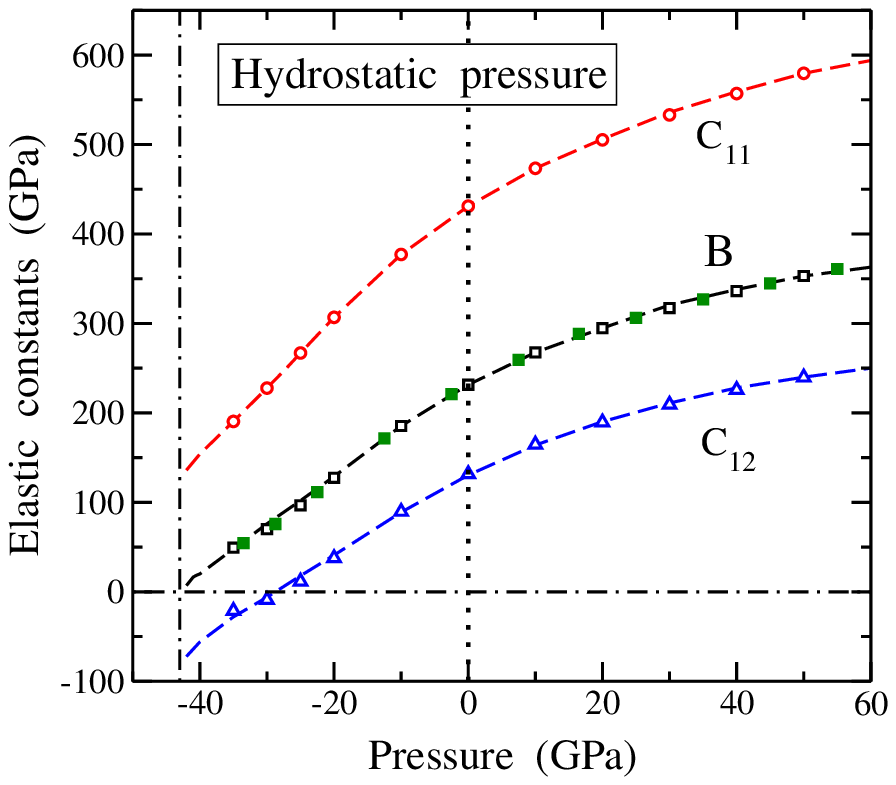}
\vspace{-5mm}
\caption{Pressure dependence of the elastic constants $C_{11}$
(circles) and $C_{12}$ (triangles) obtained from MD simulations of
$3C$-SiC at $T =$ 300~K. Open squares indicate the bulk modulus
$B$ derived from the elastic constants by using Eq.~(\ref{bulkm}).
Solid squares represent values of $B$ calculated from a numerical
derivative of the $P-V$ equation of state.
Error bars are on the order of the symbol size.
The vertical dashed-dotted line shows the spinodal pressure
at 300~K: $P_s = -43$~GPa. Dashed lines are guides to the eye.
}
\label{f6}
\end{figure}

We now address the effect of pressure on the elastic constants and its 
implications for the bulk modulus $B$ and the mechanical stability of 
the material. In Fig.~6, we show the pressure dependence of $C_{11}$ and 
$C_{12}$ under hydrostatic conditions at $T = 300$~K. Open symbols 
represent data obtained from MD simulations. 
Under compressive stress ($P > 0$), both stiffness constants increase. 
In contrast, under tensile stress ($P < 0$), they decrease, with 
$C_{12}$ becoming negative at $P = -27$~GPa. This implies that the 
Poisson's ratio, $\nu = C_{12} / C_{11}$, also becomes negative at 
this pressure, indicating that cubic SiC exhibits auxetic behavior. 
Notably, $C_{11}$ remains positive throughout the entire pressure range 
considered, satisfying a key criterion for mechanical stability in 
crystalline solids \cite{sc-ja14,mo14}.

For cubic crystals, the isothermal bulk modulus $B$ can be computed 
from the elastic constants using the relation \cite{ki05,sc-ja14}:
\begin{equation}
    B = \frac{1}{3} (C_{11} + 2 \, C_{12}) ; .
\label{bulkm}
\end{equation}
Using the $C_{11}$ and $C_{12}$ values reported above, we obtain at 
$T = 300$~K and $P = 0$ a bulk modulus of 231(1)~GPa, which falls within 
the range of experimental values reported in the literature, from 
224~GPa \cite{sc-ye71} to 260~GPa \cite{sc-yo93}. When nuclear quantum 
effects are included, our PIMD simulations yield a bulk modulus of 
$B = 224(1)$~GPa at the same temperature.

In Fig.~6, we show the pressure dependence of the bulk modulus $B$ at 300~K, 
alongside the elastic constants. Open squares represent values of $B$ 
calculated from the elastic constants using Eq.~(\ref{bulkm}). 
For comparison, we also include values obtained from a numerical derivative 
of the pressure-volume equation of state at this temperature, using the 
definition $B = -V \, \partial P / \partial V$ (solid squares). 
The close agreement between the two methods provides a robust consistency 
check for the accuracy of our calculations.

In Sec.~III.A, we inferred a spinodal pressure of $P_s = -43$~GPa at 300~K 
from the divergence in the slope of the energy-pressure curve. This value 
of $P_s$ is consistent with an extrapolation of the bulk modulus to zero 
under hydrostatic tension at the same temperature, indicating that the 
solid becomes mechanically unstable at that pressure.

\subsection{Pressure effects on the electronic band gap}

In this section, we analyze the evolution of the direct 
band gap, $E_{\Gamma}$, under hydrostatic and uniaxial pressure, 
with a particular focus on its behavior near mechanical instabilities. 
Special attention is given to the correlation between external 
pressure, the associated volume changes, and the closure of the 
band gap at instability points.

Electron-phonon interaction leads to a reduction in interband transition 
energies as temperature increases. A similar decrease occurs at low 
temperatures due to the zero-point motion of atomic nuclei.
In this context, band-gap renormalization in semiconductors due to 
electron-phonon coupling has been analyzed by examining 
changes in experimental excitation spectra over a broad temperature 
range \cite{sc-ca05,sc-ca05b}, as well as through perturbation theory
\cite{sc-zo92,sc-ki89}.

For the minimum-energy configuration of cubic SiC, the TB 
Hamiltonian predicts a direct band gap of 
$E_{\Gamma} = 7.57$~eV for the $\Gamma_{15}^v \to \Gamma_1^c$ 
transition, consistent with previous findings \cite{ra08}. 
In comparison, Kohn-Sham energy differences determined within 
the DFT framework yield a lower value of
6.18~eV. The direct band gap is a well-characterized feature of 
the electronic structure and serves as a key parameter in our 
simulations, both classical and PIMD.

\subsubsection{Hydrostatic pressure}

\begin{figure}
\vspace{-7mm}
\includegraphics[width= 7cm]{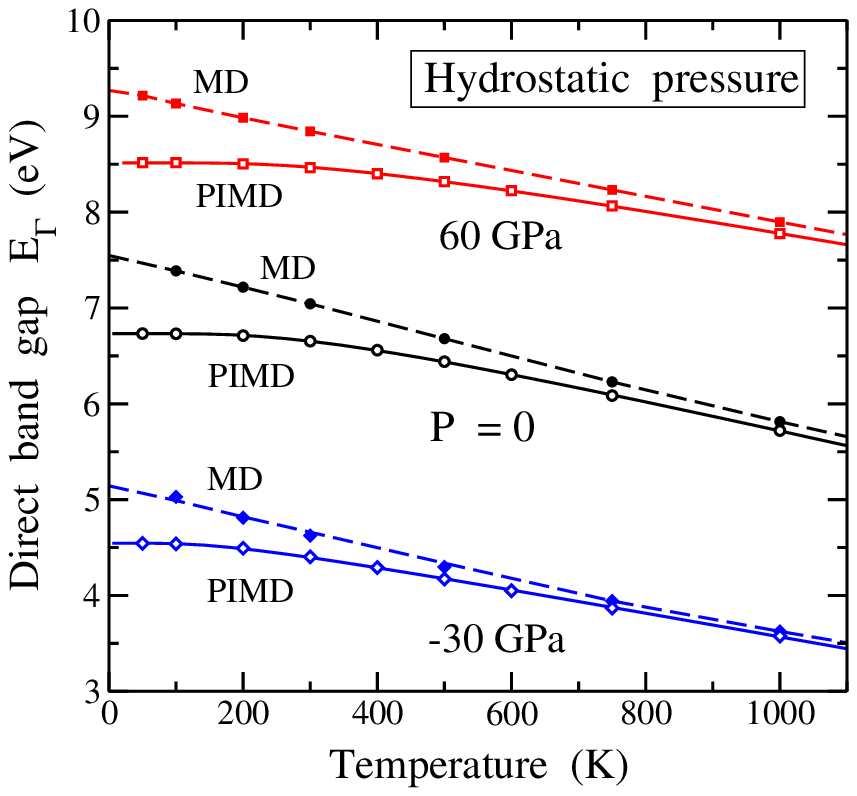}
\vspace{-5mm}
\caption{Temperature dependence of the direct band gap,
$E_{\Gamma}$. Closed and open symbols represent outcomes of
classical MD and PIMD simulations, respectively, for $P = 60$
(squares), 0 (circles), and $-30$~GPa (diamonds).
Error bars are on the order of the symbol size.
Continuous lines are fits of PIMD data to Eq.~(\ref{egt}).
Dashed lines are guides to the eye.
}
\label{f7}
\end{figure}

Fig.~7 shows the temperature dependence of the direct band gap, 
$E_{\Gamma}$, in $3C$-SiC. We present results from both classical MD 
(solid symbols) and PIMD simulations (open symbols) for hydrostatic 
pressures of $P = 0$ (circles), 60~GPa (squares), and $-30$~GPa 
(diamonds).
In all cases, $E_{\Gamma}$ decreases with increasing temperature, 
consistent with previous classical and PIMD simulations of $3C$-SiC 
at $P = 0$ \cite{ra08}. In the classical zero-temperature limit, 
$E_{\Gamma}$ increases under compression, reaching 9.32~eV at 
$P = 60$~GPa. Conversely, under tensile pressure , the direct band gap 
decreases, reaching 5.15~eV at $P = -30$~GPa.

Additionally, we observe a reduction in $E_{\Gamma}$ due to quantum 
nuclear motion across the entire temperature range shown in Fig.~7. 
This effect arises from band-gap renormalization associated with 
electron-phonon interaction.
Extrapolation of the simulation data indicates a zero-point decrease 
in $E_{\Gamma}$ of approximately 0.8~eV for both $P = 0$ and 60~GPa, 
while for $P = -30$~GPa, the reduction is lower at 0.60(2)~eV 
compared to the classical result.
In the low-temperature limit, the MSF $(\Delta r)^2$ of C
atoms is found to be $8.1$, $6.3$, and 
$5.8 \times 10^{-3}$~\AA$^2$ for $P = -30$, 0, and 60~GPa, respectively, 
indicating a significantly larger amplitude of atomic zero-point 
vibrations under tensile stress. A similar trend is observed for 
the smaller MSF of Si atoms. This behavior is linked to the overall 
decrease in the mean vibrational frequency $\overline{\omega}$ under 
tension, which reduces the impact of zero-point renormalization on 
$E_{\Gamma}$ due to electron-phonon coupling.


\begin{table*}[ht]
\caption{Direct band gap $E_{\Gamma}$ of $3C$-SiC derived from
PIMD simulations at $T$ = 0, 300 and 1000~K for hydrostatic pressure
$P = -30$, 0, and 60~GPa.
Zero-temperature data and values of $\Theta_P$ were
obtained from finite-$T$ results, using Eq.~(\ref{egt}).
Results for the Debeye temperature $\Theta_D$ were derived from
Eq.~(\ref{omd}).
Statistical error bars in $E_{\Gamma}$ are estimated to be
$\pm 0.02$~eV.
}
\vspace{0.6cm}

\begingroup
\setlength{\tabcolsep}{10pt}
\renewcommand{\arraystretch}{1.5}
\centering
\setlength{\tabcolsep}{10pt}
\begin{tabular}{|c|c c c|c c|}
\cline{2-4}
\multicolumn{1}{c}{} &
\multicolumn{3}{|c|}{$E_{\Gamma}$ (eV)}   \\ [2mm]
\cline{1-6}
 $P$ (GPa) & $T = 0$ & 300~K  & 1000~K &  $\Theta_P$ (K) & $\Theta_D$ (K) \\[2mm]
\hline
        $-30$ &  4.54  &  4.40   &  3.58   &  491  &  1114   \\
           0  &  6.73  &  6.65   &  5.72   &  894  &  1155  \\
          60  &  8.52  &  8.46   &  7.78   & 1012  &  1227  \\
\hline
\end{tabular}
\endgroup
\label{gap_simul}
\end{table*}

For each applied pressure $P$, the difference between classical MD 
and PIMD results remains noticeable over a wide temperature range 
in Fig.~7. At $T = 750$~K, this discrepancy is still evident for 
$P = 0$ and 60~GPa, whereas for $P = -30$~GPa, it falls within 
the error bars of the data points.
The solid lines in Fig.~7 represent fits of the PIMD results to a 
Bose-Einstein-type expression \cite{sc-pe98,ra08}:
\begin{equation} 
    E_{\Gamma}(P,T) = E_{\Gamma}^0(P) + \Delta_{\Gamma}(P) \; 
     \left[ 1 + \frac{2} {\exp(\Theta_P / T) - 1} \right] \; , 
\label{egt} 
\end{equation}
where $\Theta_P$ is a pressure-dependent effective temperature, and 
$\Delta_{\Gamma}(P)$ represents a negative gap renormalization 
at $T = 0$, such that
$E_{\Gamma}(P,0) = E_{\Gamma}^0(P) + \Delta_{\Gamma}(P)$.
We obtain $\Theta_P$ values of 1012, 894, and 491~K for $P = 60$, 0, 
and $-30$~GPa, respectively. This trend indicates that under tensile 
stress, the quantum data converge more rapidly to the classical limit 
as temperature increases, while under compression, this convergence 
is slower.
Results of PIMD simulations for $E_{\Gamma}$ at selected temperatures 
and hydrostatic pressures are summarized in Table~II. The faster 
approach to the classical limit under tension is consistent with 
the lower values of both $\Theta_P$ and the Debye 
temperature $\Theta_D$.

The behavior of the direct band gap $E_{\Gamma}$ is linked 
to the variation in the mean phonon frequency $\overline{\omega}$, 
which decreases under tensile stress and increases under compression, 
as discussed in Sec.~III.A. Consequently, at a given temperature $T$, 
the average thermal population of excited phonon states is higher 
under tension and lower under compression.
This trend is partially related to changes in the Debye temperature 
$\Theta_D$, which increases for $P > 0$ and decreases for $P < 0$.

\begin{figure}
\vspace{-7mm}
\includegraphics[width= 7cm]{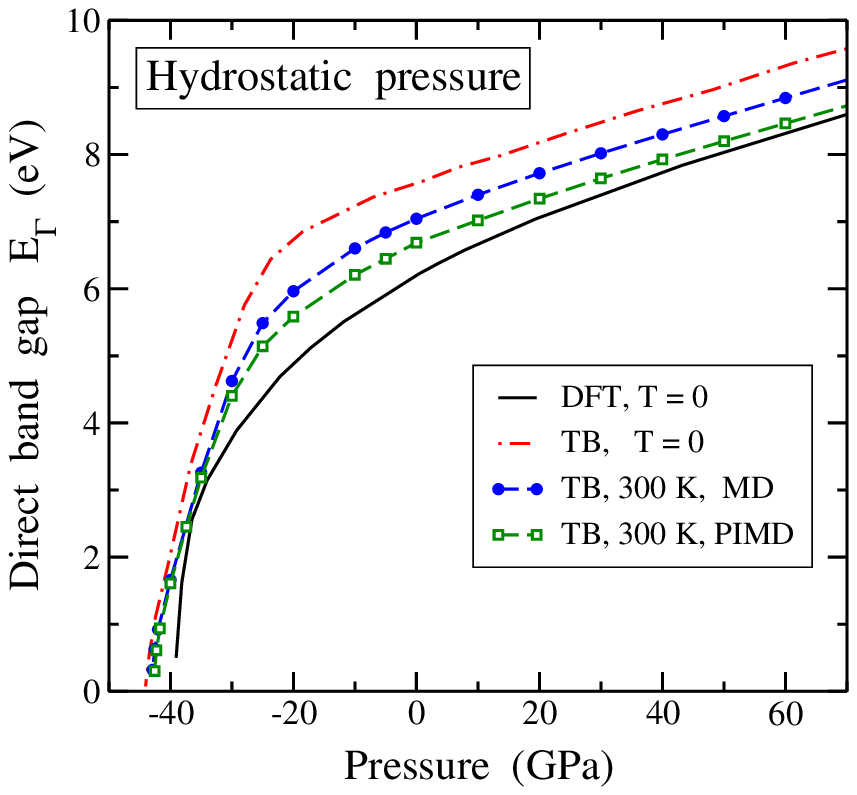}
\vspace{-5mm}
\caption{Pressure dependence of the direct band gap,
$E_{\Gamma}$. Solid circles and open squares indicate results
of classical MD and PIMD simulations, respectively, at
$T$ = 300~K.
Dashed lines through the data points are guides to the eye.
The solid and dashed-dotted lines correspond to DFT and TB
calculations at $T = 0$, respectively.
}
\label{f8}
\end{figure}

Values of $\Theta_D$ for $P = -30$, 0, and 60~GPa, obtained from 
Eq.(\ref{omd}), are listed in Table~II. These results confirm that 
$\partial \Theta_D / \partial P > 0$. However, this pressure 
dependence is significantly weaker than that of $\Theta_P$.
While variations in $\Theta_D$ contribute to changes in $\Theta_P$, 
the latter is also influenced by anharmonic effects, which 
become more pronounced with increasing temperature. Moreover, 
$\Theta_P$ is strongly affected by electron-phonon interactions, 
which themselves depend on pressure \cite{sc-ca04,sc-ca05}.

In Fig.~8, we present the pressure dependence of the direct band gap 
$E_{\Gamma}$ over a broad range, encompassing both tensile 
and compressive regimes. The solid and dash-dotted lines represent 
the DFT and TB results at $T = 0$, respectively. Symbols denote 
the direct gap obtained from classical MD (solid circles) and 
PIMD simulations (open squares) at $T = 300$~K.
All four sets of results exhibit a similar trend, showing a rapid 
decrease for $P < 0$ (tension), ultimately approaching gap closure 
near the corresponding spinodal pressure $P_s$.

Values of $E_{\Gamma}$ derived from quantum simulations appear 
lower than those from classical simulations for $P > -30$~GPa. 
Notably, under compression, the discrepancy between the two data 
sets remains approximately 0.37~eV up to 60~GPa.
In contrast, for tensile pressure ($P < -30$~GPa), the two data 
ensembles become nearly indistinguishable within the pressure range 
where $\partial E_{\Gamma} / \partial P$ increases sharply. 
Both sets extrapolate to $E_{\Gamma} = 0$ at 
$P = -44(1)$~GPa, which, within error bars, coincides with the 
spinodal pressure $P_s$ obtained from the divergence of the 
energy $E$ at $T = 300$~K (see Sec.~III.A and Fig.~2).

\begin{figure}
\vspace{-4mm}
\includegraphics[width= 7cm]{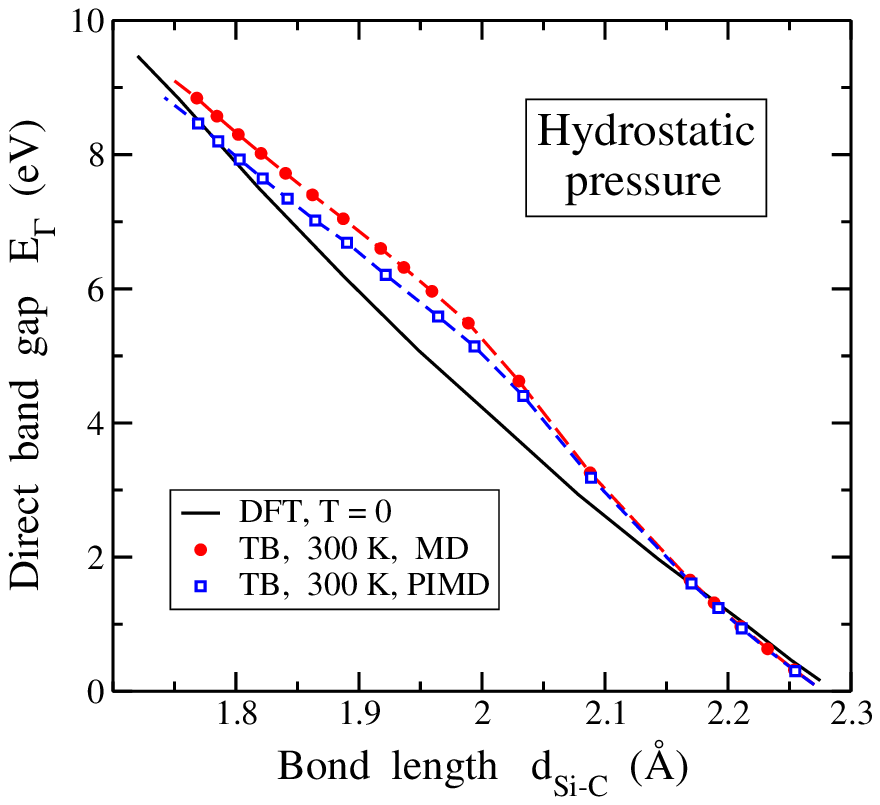}
\vspace{-5mm}
\caption{Direct band gap $E_{\Gamma}$ of $3C$-SiC as a function
of the bond length $d_{\rm Si-C}$. Solid and open symbols represent
results of classical MD and PIMD simulations, respectively,
at $T$ = 300~K. The solid line represents the DFT outcome
at $T = 0$. Lines connecting the data points are guides to the eye.
}
\label{f9}
\end{figure}

The direct band gap $E_{\Gamma}$ is closely linked to the bond 
distance $d_{\rm Si-C}$. This relationship can be understood within 
a tight-binding framework, where the evolution of $s$ and $p$ 
atomic orbitals into the valence and conduction bands at the 
$\Gamma$ point leads to the opening of the electronic gap as 
nearest-neighbor atoms move closer together \cite{yu96,sc-co17}.
Fig.~9 illustrates the dependence of $E_{\Gamma}$ on 
$d_{\rm Si-C}$. The solid line represents the DFT results at 
$T = 0$, while symbols denote values obtained from classical 
TB-MD (solid circles) and TB-PIMD simulations (open squares) at 
$T = 300$~K.
The TB-based simulations closely follow the DFT calculations for 
both large and short bond lengths. However, in the intermediate 
range, particularly at the equilibrium bond distance of the crystal 
($d_{\rm Si-C} = 1.89$~\AA), the TB method predicts a higher value 
of $E_{\Gamma}$. Additionally, for $d_{\rm Si-C} < 2$~\AA, 
the PIMD results are consistently lower than those from classical 
MD simulations, though both approaches converge at larger bond 
lengths (stronger tension). This convergence mirrors the behavior 
observed for the energy $E$ near the spinodal pressure $P_s$, 
as shown in Fig.~2.

The direct band gap $E_{\Gamma}$ obtained from our DFT calculations 
follows a trend that can be fitted to the function:
\begin{equation}
     E_{\Gamma} = E_{\Gamma}^0 + c_1 \, \delta d + 
	  c_2 \, (\delta d)^2 \; ,
\label{egg}
\end{equation}
where $\delta d = d_{\rm Si-C} - d_{\rm Si-C}^0$, with fitting 
parameters $c_1 = -18.45$~eV/\AA\ and $c_2 = 7.53$~eV/\AA$^2$. 
This equation predicts a vanishing of the direct band gap 
($E_{\Gamma} = 0$) 
at $d_{\rm Si-C} = 2.28(1)$~\AA, corresponding to a lattice 
parameter of 5.26~\AA.
Similarly, extrapolation of TB data at $T = 300$~K yields 
$E_{\Gamma} = 0$ at $d_{\rm Si-C} = 2.27(1)$~\AA, closely 
matching the DFT result at $T = 0$.

Near $P_s$, the direct band gap can be expressed through a 
first-order expansion as $E_{\Gamma} = \gamma (a_s - a)$ for 
$a < a_s$, where $\gamma$ is a positive constant that may depend 
on the specific model under consideration. Substituting in
Eq.~(\ref{asa}), we obtain
\begin{equation}
   E_{\Gamma} = \frac{c \gamma}{3 a_s^2} \, (P - P_s)^{1/2} \; ,
\label{egg1}
\end{equation}
which aligns with the behavior observed in Fig.~8 close to $P_s$.
The relationship between the constant $\gamma$ and the parameters 
$c_1$ and $c_2$, which describe the dependence of $E_{\Gamma}$ 
on the bond length, is provided in Appendix~B.

Our DFT calculations predict an indirect band gap of
$E_g = 1.24$~eV ($\Gamma_{15}^v \to X_1^c$) for the minimum-energy 
configuration of $3C$-SiC with a lattice parameter of $a = 4.36$~\AA. 
Under hydrostatic pressure, the behavior of $E_g$ contrasts with that 
of the direct band gap $E_{\Gamma}$: $E_g$ decreases with increasing 
compression, indicating that $\partial E_g / \partial P < 0$ 
(see Refs.\cite{sc-gh89,sc-fr98,sc-we99}). For instance, when 
$a = 3.6$~\AA, $E_g$ drops to 0.56~eV. Conversely, under tensile 
strain, $E_g$ increases, reaching 1.48~eV at $P = -38$~GPa 
($a = 5.1$~\AA), near the spinodal pressure $P_s$.
While our DFT calculations underestimate the indirect band gap compared 
to experimental values of approximately 2.2--2.4~eV for cubic SiC
\cite{sc-ch64,sc-mo94,sc-lb82}, they successfully capture the overall 
pressure-dependent trends. Notably, under tensile strain, the direct band gap 
$E_{\Gamma}$ decreases and eventually becomes the smallest or fundamental
electronic band gap. These results indicate that $3C$-SiC transitions 
to a direct-gap semiconductor at $a \approx 5$~\AA, in the vicinity of 
the spinodal pressure $P_s$.

\subsubsection{Uniaxial pressure}

\begin{figure}
\vspace{-7mm}
\includegraphics[width= 7cm]{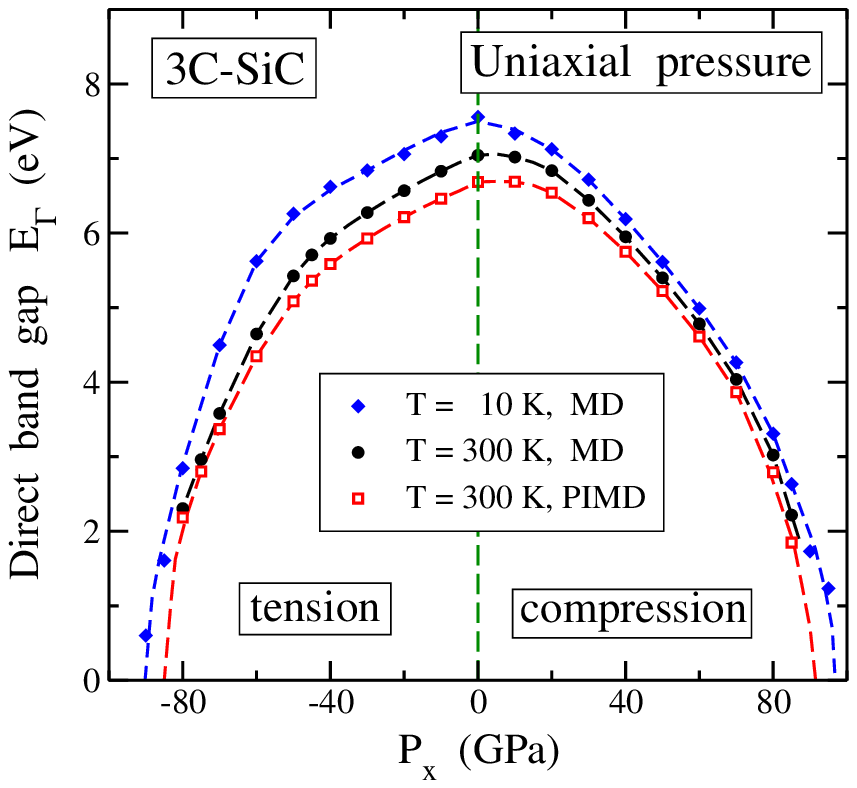}
\vspace{-5mm}
\caption{Direct band gap $E_{\Gamma}$ as a function of
uniaxial pressure $P_x$.
Symbols are data points obtained from finite-temperature
simulations: classical at
$T = 10$~K (solid diamonds), classical at
$T = 300$~K (solid circles), and PIMD at $T = 300$~K
(open squares).
Error bars are of the order of the symbol size.
Dashed lines are guides to the eye.
}
\label{f10}
\end{figure}

We now extend the analysis of the direct band gap $E_{\Gamma}$, 
previously determined under hydrostatic pressure, to examine 
the effects of uniaxial pressure applied along one of the axes of 
the cubic unit cell of $3C$-SiC, specifically the $x$-axis, 
where $P_x = -\sigma_{xx}$. In Fig.~10, we present the dependence 
of $E_{\Gamma}$ on the uniaxial pressure $P_x$. The results from 
our classical MD simulations at $T = 10$~K (solid diamonds) 
and $T = 300$~K (solid circles) are shown, alongside those 
from PIMD simulations at $T = 300$~K (open squares).

For tensile uniaxial pressure ($P_x < 0$), Fig.~10 shows a 
reduction of the direct band gap with increasing tension, similar 
to the behavior observed under hydrostatic pressure for $P < 0$. 
However, in this case, $E_{\Gamma}$ reaches a maximum at 
$P_x \approx 0$ before decreasing further under increasing 
compression, eventually reaching a pressure where the direct band
gap vanishes. This trend mirrors the anomalies in the lattice 
parameter and bond length observed near $-84$~GPa and 90~GPa, as 
shown in Figs.~4(a) and 5(a).
By considering a low temperature ($T = 10$~K), we can approach 
the instability limits more closely than at higher temperatures, 
such as $300$~K. As $T$ increases, fluctuations in the simulation 
cell volume become more significant, leading to the failure of 
our $NPT$ simulations before reaching the instability pressure, 
which coincides with the point at which the Young's modulus 
$Y$ vanishes.

Near the tensile pressure $P_x^{i1}$, the direct band gap
$E_{\Gamma}$ follows a dependence analogous to that observed 
under hydrostatic pressure (see Eq.~(\ref{egg1})):
\begin{equation}
   E_{\Gamma} = A_1 \, (P_x - P_x^{i1})^{1/2} \; ,
\label{ega1}
\end{equation}
and similarly, near $P_x^{i2}$, we have
\begin{equation}
   E_{\Gamma} = A_2 \, (P_x^{i2} - P_x)^{1/2} \; ,
\label{ega2}
\end{equation}
where $A_1$ and $A_2$ are constants.
Fitting our classical simulation results at $T = 300$~K to 
Eqs.~(\ref{ega1}) and (\ref{ega2}), using the four data points 
closest to the instability pressure in each case, we obtain 
$P_x^{i1} = -85(2)$~GPa and $P_x^{i2} = 92(2)$~GPa, in 
agreement with the results presented in Secs.~III.B and III.C. 
A similar fitting for the $T = 10$~K data yields 
$P_x^{i1} = -90(1)$~GPa and $P_x^{i2} = 97(2)$GPa, 
indicating a shift of approximately 5~GPa in the instability 
pressures at lower temperature, thereby expanding the stability 
region.  Notably, extrapolation of the PIMD data shown in 
Fig.~10 yields instability pressures that, within error bars, 
are indistinguishable from those derived from classical 
simulations.

The hydrostatic component of an applied stress 
$\{ \sigma_{ij} \}$ is given by
\begin{equation}
	P_H = -\frac{1}{3} {\rm Tr}({\bf \sigma}) =
   -\frac{1}{3} (\sigma_{xx} + \sigma_{yy} + \sigma_{zz}) \; ,
\label{ph13}
\end{equation}
where ${\rm Tr}(\sigma)$ denotes the trace of the stress tensor.
In the present case, we have $P_H = P_x / 3$, implying that for 
the instability pressures $P_x^{i1}$ and $P_x^{i2}$, the 
corresponding hydrostatic components are $P_H = -28$~GPa and 
$P_H = 30$~GPa, respectively.
Notably, the former value is lower than the tensile spinodal 
pressure: $P_s = -43$~GPa.

The vanishing of the direct band gap $E_{\Gamma}$ at the uniaxial 
pressures $P_x^{i1}$ and $P_x^{i2}$ is directly linked to 
the vanishing of the Young's modulus ($Y \to 0$), signaling 
the onset of mechanical instability in the crystal structure. 
This instability is reflected in the anomalies observed in the 
lattice parameter and bond length (Figs.~4 and 5).
In this context, the vanishing of the Young's modulus 
corresponds to the divergence of the derivative 
$\partial a_x / \partial P_x$, as described in Eq.~(\ref{young}).

At the instability pressures $P_x^{i1}$ and $P_x^{i2}$, the 
bond length $d_{\rm Si-C}$ converges to approximately 2.10~\AA\ 
and 1.93~\AA, respectively (see Fig.~5(a)). The former, 
corresponding to tensile stress, is notably shorter than the 
bond length obtained at the spinodal point $P_s$ under 
hydrostatic pressure ($d_{\rm Si-C} = 2.27$~\AA).
This indicates that the direct band gap collapses under uniaxial 
pressure (both tensile and compressive) at significantly shorter 
interatomic distances than under hydrostatic pressure, where 
the cubic crystal structure remains intact.
In this context, deviations from cubic symmetry can be 
quantified by the ``tetragonality factor,'' defined as the 
ratio $a_{\perp} / a_x$ under uniaxial pressure $P_x$. 
At $T = 300$~K, this ratio varies substantially, ranging 
from 0.72 at $P_x^{i1}$ (tension) to 1.68 at $P_x^{i2}$ 
(compression), as inferred from the data presented 
in Fig.~4(a).

\begin{figure}
\vspace{-7mm}
\includegraphics[width= 7cm]{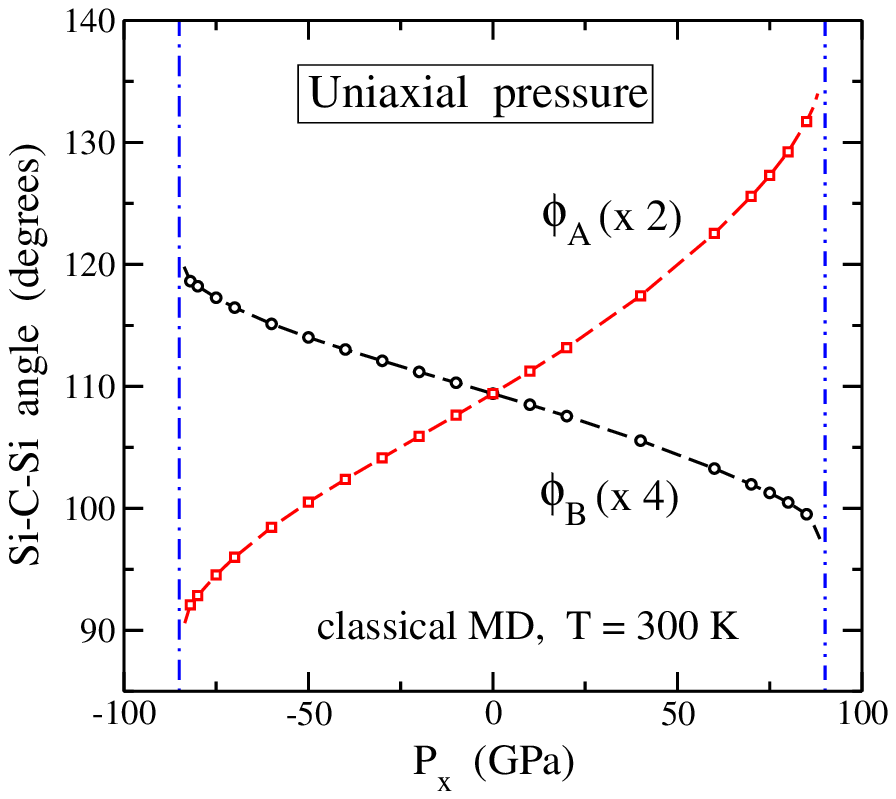}
\vspace{-5mm}
\caption{Si-C-Si angle vs uniaxial pressure $P_x$.
Open symbols are data points derived from our
classical MD simualtions at 300~K: squares for $\phi_A$
(multiplicity 2) and circles for $\phi_B$ (multiplicity 4).
Error bars are less than the symbol size.
Dashed lines are guides to the eye.
Dashed-dotted lines indicate the instability pressures
$P_x^{i1} < 0$ and $P_x^{i2} > 0$.
}
\label{f11}
\end{figure}

Under uniaxial pressure, the electronic structure is influenced 
not only by changes in bond length but also by modifications 
in the structural C-Si-C and Si-C-Si bond angles. Notably, 
under an applied pressure $P_x$, the six C-Si-C angles formed 
by the four carbon atoms bonded to a silicon atom undergo distinct 
changes: two angles increase under compression (denoted as $\phi_A$ 
in Fig.~11), while the remaining four decrease (labeled as 
$\phi_B$). A similar behavior is observed for the Si-C-Si angles 
centered at a carbon atom.
These variations highlight the significant distortions experienced 
by the initial tetrahedral geometry under the applied pressure $P_x$. 
Near the instability pressures $P_x^{i1}$ and $P_x^{i2}$, we observe 
a sharp increase in the slope $\partial \phi_A / \partial P_x$, 
resembling the steep pressure dependence of $a_x$ shown in Fig.~4(a). 
In contrast, the variation of $\partial \phi_B / \partial P_x$ in 
these regions is less pronounced.
These angle modifications, combined with changes in bond length, 
result in substantial variations in next-nearest-neighbor distances 
(C--C and Si--Si). At $T = 300$~K, these distances range from 
3.6 ~\AA\ near $P_x^{i1}$ to 2.9~\AA\ at $P_x^{i2}$. For comparison, 
the equilibrium unstressed structure exhibits a next-nearest-neighbor 
distance of 3.08~\AA.

\section{Summary}

In this work we have carried out a systematic computational study of 
$3C$-SiC under hydrostatic as well as uniaxial pressure, covering 
the pressure range in which the crystal is found to be mechanically 
stable, both in the compressive and tensile regimes. 
We have employed a well-tested TB model to describe the system at
finite temperatures, performing simulations in the $NPT$ ensemble, 
incorporating quantum effects within the PIMD formalism and in 
the classical limit, so as to gauge the relevance of quantum effects 
in the studied properties. The validity of the TB model has been 
checked against first-principles zero temperature calculations 
employing DFT. We have thoroughly characterised the behavior of the
system total energy, its structural properties and the electronic 
band gap at finite temperatures and tensile/compressive pressures.

Our analysis examines structural properties such as the lattice 
parameter, Si--C bond length, and bond angles, along with their 
fluctuations as functions of pressure and temperature. 
The onset of mechanical instability is investigated by evaluating 
the bulk modulus $B$ under hydrostatic pressure and Young's modulus 
$Y$ under uniaxial pressure.

We have identified instabilities in the silicon carbide structure 
under specific pressure conditions. At $T = 300$~K, the bulk 
modulus vanishes under hydrostatic pressure at $P_s = -43$~GPa, 
indicating a divergence in compressibility at the spinodal point. 
Under uniaxial pressure along the [100] crystal axis, the Young's 
modulus approaches zero ($Y \to 0$) for a tensile pressure of 
$P_x^{i1} = -84$~GPa and a compressive one of 
$P_x^{i2} = 90$~GPa. These pressures define the mechanical stability 
limits of the material. When these limits are reached, several 
properties exhibit anomalies, with some, such as the lattice 
parameter and its MSF, showing divergences.

The direct band gap $E_{\Gamma}$ decreases significantly with 
increasing temperature and exhibits a rapid reduction under 
tensile strain for both hydrostatic and uniaxial pressure. 
However, under compression, the pressure derivative behaves 
differently: $\partial E_{\Gamma} / \partial P > 0$ for 
hydrostatic pressure, while 
$\partial E_{\Gamma} / \partial P_x < 0$ under uniaxial pressure 
$P_x$. In all cases, $E_{\Gamma}$ vanishes under 
pressure conditions that coincide with the loss of 
mechanical stability.

While classical MD simulations provide reasonable approximations 
for certain properties, such as lattice parameters and bond 
lengths, other properties are significantly affected by nuclear 
quantum effects. Notably, at low temperatures, the direct band gap 
$E_{\Gamma}$ exhibits pronounced quantum contributions due to 
electron-phonon coupling, leading to a noticeable renormalization 
of the gap, as shown in Fig.~7. Nuclear quantum motion also 
manifests in the fluctuations of structural variables, such as 
bond length, as illustrated in Fig.~5(b). However, it is 
important to note that these quantum effects do not significantly 
alter the instability pressures $P_s$, $P_x^{i1}$, and $P_x^{i2}$ 
obtained from classical MD simulations.

Atomistic simulations similar to those presented here can also 
offer valuable insights into the behavior of related materials, 
particularly tetrahedral semiconductors, under tensile and 
compressive stress. Furthermore, quantum nuclear motion can 
significantly influence various properties of these materials 
at relatively low temperatures. Such effects can be investigated 
through atomic-scale simulations using techniques like the PIMD 
method employed in this study.

\begin{acknowledgments}
This work was supported by Ministerio de Ciencia e Innovaci\'on
(Spain) through Grant PID2022-139776NB-C66.  \\
\end{acknowledgments}

\noindent
{\bf Data availability} \\

The data that support the findings of this article are openly 
available \cite{sc-ze25b}.  \\ \\


\appendix

\section{Debye model}

In the Debye model, the vibrational spectrum of a solid is
approximated by a VDOS given by $\rho_D(\omega) = C \omega^2$,
which extends up to a maximum value $\omega_D$, and $C$ is
a normalization constant to account for the vibrational
degrees of freedom in the material. Using the same 
normalization as that given in Eq.~(\ref{intg}), we have
\begin{equation}
   \int_0^{\omega_D} \rho_D(\omega) \, d\omega = 6 \; .
\end{equation}
This corresponds to a crystallographic cell including
two atoms, so we have: $C = 18 / \omega_D^3$.

In this approximation, the vibrational energy per atom 
at temperature $T$ is given by:
\begin{equation}
 E(T) = E_{\rm ZP} + \frac12 \int_0^{\omega_D}
   n(\omega) \, \hbar \omega \, \rho_D(\omega) \, d \omega \; ,
\label{ete}
\end{equation}
where $n(\omega)$ is the Bose-Einstein factor:
\begin{equation}
   n(\omega) = \left[ {\rm e}^{\beta \hbar \omega} - 1 
	\right]^{-1} \; ,
\label{nom}
\end{equation}
with $\beta = 1 / k_B T$ ($k_B$, Boltzmann constant).
The zero-point energy per atom is given by:
\begin{equation}
 E_{\rm ZP} = \frac12 \int_0^{\omega_D} \frac{\hbar \omega}{2} \,
      \rho_D(\omega) \, d \omega  \; .
\label{ezp2}
\end{equation}
Note that the prefactor $1/2$ in Eqs.~(\ref{ete}) and (\ref{ezp2}) 
gives the energy per atom.
From Eq.~(\ref{ezp2}), we find
\begin{equation}
      E_{\rm ZP} =  \frac98 \hbar \, \omega_D =
	  \frac98  k_B \Theta_D  \; ,
\label{ezp3}
\end{equation}
which gives us a relation between the zero-point energy and
the Debye temperature $\Theta_D$.

\section{Fitting of the energy gap}

We take the DFT results as a reference for the dependence 
of the direct band gap $E_{\Gamma}$ on bond length under 
isotropic volume changes (or hydrostatic pressure).
Using a Taylor expansion for $E_{\Gamma}$ around the 
equilibrium bond length, $d_{\rm Si-C}^0$, we have: 
\begin{equation}
  E_{\Gamma} = E_{\Gamma}^0 + 
	c_1 \, (d_{\rm Si-C} - d_{\rm Si-C}^0) +
         c_2 \, (d_{\rm Si-C} - d_{\rm Si-C}^0)^2  \; ,
\label{egg3}
\end{equation}
where $c_1$, $c_2$ are fitting constants.

Near the spinodal point, the dependence of $E_{\Gamma}$ on the
lattice parameter may be written for $a < a_s$ as:
\begin{equation}
	E_{\Gamma} =  \gamma (a_s - a)
\label{egg4}
\end{equation}
where $a_s$ is the lattice parameter for the spinodal 
pressure $P_s$.
Taking into account that we have for $3C$-SiC the relation
$a = k \, d_{\rm Si-C}$, with $k = 4 / \sqrt{3}$, then
\begin{equation}
   \frac{\partial E_{\Gamma}}{\partial a } = \frac{c_1}{k} 
	+ 2 \, \frac{c_2}{k^2} \, (a - a_0)  \; .
\label{deg}
\end{equation}
Thus,
\begin{equation}
 \gamma = - \left. \frac{\partial E_{\Gamma}}{\partial a} 
    \right|_{a_s} =
   2 \, \frac{c_2}{k^2} \, (a_0 - a_s) - \frac{c_1}{k}  \; .
\label{egg5}
\end{equation}
This provides us with a relation between the slope of the
$a$--$E_{\Gamma}$ curve at the spinodal point and the
parameters $c_1$ and $c_2$ describing the variation of
$E_{\Gamma}$ around the equilibrium structure.



%

\end{document}